# Robust Depth Linear Error Decomposition with Double Total Variation and Nuclear Norm for Dynamic MRI Reconstruction

Junpeng Tan, Chunmei Qing, Xiangmin Xu

**Abstract.** Compressed Sensing (CS) significantly speeds up Magnetic Resonance Image (MRI) processing and achieves accurate MRI reconstruction from under-sampled k-space data. According to the current research, there are still several problems dynamic MRI k-space reconstruction based on CS. 1) There are differences between Fourier domain and Image domain, and the differences between MRI processing of different domains needs to be considered. 2) As a three-dimensional data, dynamic MRI has its spatial-temporal characteristics, which needs calculate the difference and consistency of surface textures while preserving structural integrity and uniqueness. 3) Dynamic MRI reconstruction is time-consuming and computationally resource dependent. In this paper, we propose a novel robust low-rank dynamic MRI reconstruction optimization model via highly under-sampled and Discrete Fourier Transform (DFT), which is called Robust Depth Linear Error Decomposition Model (RDLEDM). Our method main includes linear decomposition, double Total Variation (TV) and double Nuclear Norm (NN) regularizations. By adding linear image domain error analysis, the noise is reduced after under-sampled and DFT processing and the anti-interference ability of the algorithm is enhanced. Double TV and NN regularizations can utilize both spatial-temporal characteristics and explore the complementary relationship between different dimensions in dynamic MRI sequences. In addition, Due to the non-smoothness and non-convexity of TV and NN terms, it is difficult to optimize the unify objective model. To address this issue, we utilize a fast algorithm by solving a primal-dual form of the original problem. In comparison with five state-of-the-art methods, extensive experiments on dynamic MRI data demonstrate the superior performance of the proposed method in terms of both reconstruction accuracy and time complexity.

# 1 Introduction

Magnetic Resonance Image (MRI) has become the most mainstream clinical diagnosis and scientific research with its high resolution and non-invasive characteristics. One of the technical challenges of MRI is to reduce the acquisition time and process MRI into high spatial-temporal resolution imaging in a short time. In general, MRI reconstruction based on under-sampled is a great way to speed up MRI processing time. However, insufficient sampling will violate Nyquist sampling criteria and result in artifacts in reconstructed dynamic MRI. To overcome this problem, a promising technology known as Compressed Sensing (CS) [1] has been widely studied and used in Signal Processing [2], Inverse Problems [3], and Medical Imaging Processing [4]. CS theory shows that if a signal has a sparse representation in transform domain, it is possible to reconstruct the signal/image from fewer measurements with little or no information loss than expected in Nyquist sampling criteria.

In most cases, CS usually combines other transforms. Like, Total Variation (TV), Discrete Cosine Transform (DCT), Discrete Wavelet Transform (DWT), and Contourlet Transform (CT) as fixed transforms conducted on the image are commonly used in conventional CS-MRI reconstruction methods [5]. Therefore, many deformation optimization methods have emerged, such as Compressed MRI using Total Variation and Wavelets (TVCMRI) [6], Quantitative Analysis of Under-sampled Human Brain Sodium MRI (QAHBS) [7], the Hybrid Galactic Swarm Optimization and Grey Wolf Optimization (HGSGWO) [8]. According to these methods analysis, the above transforms play a crucial role in MRI boundary reconstruction, but some surface texture details still need to be improved. It is noteworthy that robust low-rank constraint is widely used in the field of noise reduction and is also favored by researchers in the field of MRI high quality reconstruction, such as Nuclear Norm (NN), $L_p$-norm, $L_{p,q}$-norm. Ulas et al. [9] combined TV with NN regularization to reconstruct MRI. Yang et al. [10] proposed the low-rank prior as a weighted schatten $L_p$-norm to obtain the global information of MRI. Chen et al. [11] designed $L_{1,q}$-norm sparse representation framework to preserve the edge structural and texture details. However, for dynamic MRI sequences reconstruction, since it is naturally represented as a tensor-structured data type, matrix-based methods could lead to the loss of implicit and spatial-temporal correlation structure information. Mubarak et al. [12] proposed Higher Order Dictionary Learning (HODL) for dynamic MRI reconstruction, which exploited the temporal gradient sparsity. Liu et al. [13] adapted Smooth Robust Tensor Principal Component Analysis (SRTPCA) to exploit the spatial and temporal dynamic MRI data structures. However, the methods based on these optimization models do not consider the differences between different domains, such as the existence of some missing information or redundant information after the transformation of under-sampled data after DFT processing. We need to do some depth error decomposition and preserve more structure information.

On the other hand, Deep learning frameworks have achieved good performance in computer vision applications such as image denoising [14], video inpainting [15], and 3D reconstruction [16]. The structure of dynamic MRI sequences based on deep learning [17-19] has also become a popular processing method. The main problem of MRI reconstruction based on deep learning is the consumption of computational resources and running time. Therefore, some accelerated deep learning methods have gradually become a research hotspot. Yoo et al [20] introduced a generalized version of the deep-image-prior approach, which optimized the weights of a reconstruction network. Zou et al. [21] proposed a generative Smoothness Regularization on Manifolds Model (SoRMM). Although the deep dynamic MRI reconstruction framework based on accelerated processing can show good reconstruction performance, it faces the need of many MRI training samples. Moreover, the source and preprocessing of medical data need to be checked by medical researchers, which has become the biggest obstacle to restricting interested scientific research enthusiasts. The realization of unsupervised dynamic MRI reconstruction has a universal development direction.

According to the above analysis, there are several problems in the study of unsupervised dynamic MRI sequences reconstruction: 1) The differences between the dynamic MRI sequences and the image domain through discrete Fourier transform under-sampled. We need to address the missing information and remove redundant information in the dynamic MRI reconstruction composition after under-sampled DFT processing; 2) The dynamic MRI has its spatial-temporal characteristics, which needs calculate the difference and consistency of surface textures while preserving structural integrity and uniqueness. We should consider the restoration of surface texture details and the identification of boundary information in dynamic MRI reconstruction under comprehensive factors; 3) Due to the non-smoothness and non-convexity of TV and NN terms, it is difficult to optimize the unified objective model, and then, dynamic MRI reconstruction is time-consuming and computationally complexity. The accelerated operation of the unified optimization model for dynamic MRI sequences reconstruction.

To complete the above challenge, in this paper, we further improve the reconstruction performance in compressed sensing of dynamic MRI. A novel smooth robust depth linear error decomposition model (RDLEDM) is proposed, which takes advantage of these two minimize decomposition module in the under-sampled DFT k-space domain and reconstruction image domain. The purpose of using two minimization optimization decomposition modules is to continuously reduce the reconstruction error of the two domains of k-space Fourier domain and Image domain, and to enhance the correlation different domains. Two domain reconstruction information achieve complement and eliminate differences in the algorithm optimization process. Moreover, the proposed algorithm adopts double TV and double NN regression terms, which not only ensures the recovery of more detailed textures and clear boundaries of dynamic MRI sequences from the under-sampled k-space, but also ensures the removal of irrelevant noises. We consider the restoration of surface texture details and the identification of boundary information in 3D MRI data under comprehensive factors. Compared with other recovery methods, the proposed method is an unsupervised learning method and does not require training data.

The main contribution can be listed as follow:

1) We utilize two decomposition models to deal with the reconstruction differences of 3D dynamic MRI in different domains. One is the under-sampled reconstruction module based on the multi-matrix multiplication in the Fourier domain, and the other is the full-sampled based on the linear error decomposition module in the Image domain.

2) We introduce double TV and double NN, both of which constrain the dynamic MRI based on under-sampled reconstruction and the dynamic MRI based on linear decomposition. Combined with the low-rank error regression mechanism, high-quality reconstruction of 3D dynamic MRI under comprehensive factors is achieved.

3) We use a fast optimization method to improve the computational efficiency of the unified

optimization model for 3D dynamic MRI. Experiments based on cardiac perfusion and cine images show that the proposed method outperforms the state-of-the-art algorithms in terms of Root Mean Square Error (RMSE) and Peak Signal-to-Noise Ratio (PSNR)

The rest parts of this paper are organized as follows. In Section 2, we give the notations of parameters and variables, and defined our proposed model on dynamic MRI sequences reconstruction. The optimization model for RDLEDM is given in Section 3. In Section 4, the detailed solution is described, and the results of numerical experiments are displayed. The conclusion is drawn in Section 5.

## 2 Rated Work

Firstly, we define dynamic MRI data as $X = \{x_1, x_2, \cdots, x_t, \cdots, x_T\} \in \mathbb{C}^{T \times m \times n}$, $x_t \in \mathbb{C}^{m \times n}$ denotes as one dynamic MR frame at time $t$. The MRI acquisition model in k-space can be defined as

$$b_{k,t} = \int_r^N x_{r,t} exp^{(-2j\pi k, r)} dr + e_{k,t} \tag{1}$$

Where $r$, $k$ and $N$ are the locations of spatial, k-space and all locations of spatial, respectively. $b_{k,t}$ expresses the observation, and $e$ is noise in k-space. Since dynamic MRI sequences are complex-valued space as $\langle A, B \rangle = tr(A^H B)$, and where $A^H$ denotes the Hermitian transpose of $A$. The Frobenius norm now is defined as $\|A\|_F = \sqrt{tr(A^H A)}$ and $\|A\|_F^2 = tr(A^H A)$. If $R_t$ is the under-sampled mask to acquire only a subset of k-space and the 2D DFT can be denoted $F$. The physical model for the under-sampled k-space measurement of $x_t$ can be formulated as [22]

$$b_t = R_t(Fx_t + e_t) \tag{2}$$

Since Eq. (2) is ill-posed and require low rank and smooth regulation, some CS-base k-space dynamic MRI reconstruction methods were proposed to exploit the temporal correlation. This problem can be defined as:

$$\min_{X} \varphi(X) + \phi(X)$$

$$s.t. \sum_{t=1}^{T}\|R_t F x_t - b_t\|_2^2 < \varepsilon \quad (3)$$

Here, $\varphi(\cdot)$ is the smooth regulation term, and $\phi(\cdot)$ is the low-rank regulation term. The variable $\varepsilon$ denotes dynamic MRI sequences reconstruction error in k-space. In some cases, NN regulation is used to sparse input sample data as low rank function, $\phi(\cdot) = \|X\|_* = tr\sqrt{X^T X}$, if we take care of the eigenvalue decomposition $X = U\Sigma V^T$, where $V^T V = I$ and $U^T U = I$. The more simplified transformation is $\|X\|_* = tr(\Sigma)$. TV regulation is better to process the sample boundary information as smooth function by directional gradient, $\varphi(\cdot) = \|X\|_{TV}$

$$\|u\|_{TV} = \int |\nabla u|\, dx\, dy$$

$$= \sum_i \sum_j (|u_{i+1,j} - u_{i,j}| + |u_{i,j+1} - u_{i,j}|) \quad (4)$$

Where $\int |\nabla u|$ represents the total variation of an image $u$, $|u_{i+1,j} - u_{i,j}| + |u_{i,j+1} - u_{i,j}|$ is the sum of the absolute values of the difference between the pixels in image $u$ and the pixels in vertical and horizontal directions. According to Ref. [23], the classical k-space dynamic MRI reconstruction with TV and NN regulations can be expressed as follows

$$\min_{X} \frac{1}{2}\|RFX - B\|_F^2 + \alpha\|X\|_{TV} + \beta\|X\|_* \quad (5)$$

where the parameters $\alpha$ and $\beta$ are trade-off the importance of TV and NN regulations. The first term in Eq. (5) is the general under-sampled k-space construction decomposition module, which $B$ denotes as the input under-sampled dynamic MRI data in the Fourier domain, $X$ is the full-sampled high-quality dynamic MRI data in the Image domain that needs to be reconstructed. The second term in Eq. (5) is TV constraint, and the third term is NN constraint.

## 3 The Unified Optimization Model

Without loss of generality, most under-sampled k-space reconstruction algorithms of dynamic MRI sequences, such as Eqs. (3) and (5), only consider the DFT k-space transform error, and do not consider reconstruction errors in the Image domain and Fourier domain. To further eliminate the redundant noise in the Image domain, in this paper, we propose a novel robust low-rank dynamic MRI sequences reconstruction model, which includes two linear decomposition modules. One is the DFT k-space transform, and other is image domain reconstruction linear error decomposition. Our purpose is to reconstruct dynamic MRI in k-space well, fully consider the correlation and difference of different spatial domains, eliminate redundant noise, and improve the visualization effect of dynamic MRI sequences. Therefore, we can give the objective function as

$$\min_{X,X',\varepsilon} \|RFX - B\|_F^2 + \|\varepsilon\|_* + \|X - X' - \varepsilon\|_F^2 \quad (6)$$

Where The variable $X' = \{x'_1, x'_2, \cdots, x'_t, \cdots, x'_T\} \in \mathbb{C}^{T \times m \times n}$, $x'_t \in \mathbb{C}^{m \times n}$ is the dynamic MRI sequences of the linear error decomposition in the Image domain. $\varepsilon = \{\varepsilon_1, \varepsilon_2, \cdots, \varepsilon_t, \cdots \varepsilon_T\} \in \mathbb{C}^{T \times m \times n}$, $\varepsilon_t \in \mathbb{C}^{m \times n}$ is the depth error variable of under-sampled k-space reconstruction dynamic MRI $X$ and the Image domain linear decomposition $X'$. By adopting low-rank constraints on the error variables, the purpose of this is to prevent over-fitting of $X$ during linear decomposition. At the same time, try to remove the redundant information after reconstruction of down-sampled K-space and make up for the shortcomings after DFT transformation.

On the other hand, the dynamic MRI has its spatial-temporal characteristics, which needs calculate the difference and consistency of surface textures while preserving structural integrity and uniqueness. We should consider the restoration of surface texture details and the identification of boundary information in dynamic MRI reconstruction under comprehensive factors. Based on double TV regulation, $\|\cdot\|_{TV}$, and NN regularization, $\|\cdot\|_*$ mechanism is used, which constrain the dynamic MRI $X$ based on under-sampled reconstruction and the dynamic MRI $X'$ based on linear decomposition. The unified optimization model of our proposed method RDLEDM can be listed as follows.

$$\min_{X,X',\varepsilon}(\|RFX - B\|_F^2 + \|\varepsilon\|_* + \lambda_1(\|X\|_{TV} + \|X'\|_{TV}) + \lambda_2(\|X\|_* + \|X'\|_*) \quad (7)$$

$$s.t. X = X' + \varepsilon$$

where the parameters $\lambda_1$ and $\lambda_2$ are trade-off the importance of TV and NN regulations. Our framework solves the inherent challenges associated with sparse and heterogeneous input data by reconstructing errors in k-space domain and Image domain and enhancing the depth distribution of correlation and spatial perception.

## 4 Algorithm Optimization

In this section, our main task is to solve variables $X, X', \varepsilon$. According to the prior knowledge of TV and NN, since optimization problems of both TV and NN regulations are non-smooth and non-convexity, it is difficult to effectively solve the following problem (7) through general optimization process methods.

$$\min_{X,X',\varepsilon} \|RFX - B\|_F^2 + \|\varepsilon\|_* + \|\varepsilon - X' - X\|_F^2 + \langle \tau, \varepsilon - X' - X \rangle \quad (8)$$

$$+ \lambda_1(\|X\|_{TV} + \|X'\|_{TV}) + \lambda_2(\|X\|_* + \|X'\|_*)$$

The Legendree-Fenchel transformation with total variation is used [1] [23], then the prime-dual form of the problem (8) can be obtained

$$\min_{X,X',\varepsilon} \max_Y \frac{1}{2}\|\mathcal{A}X - B\|_F^2 + \lambda_2(\|X\|_* + \|X'\|_*) + \lambda_1 \mathcal{R}\{\nabla X, Y\}$$

$$+ \lambda_1 \mathcal{R}\{\nabla X', Y\} - I_{B_\infty}(Y) + \|\varepsilon\|_* + \tau\|\varepsilon - X' - X\|_F^2 \quad (9)$$

Where $Y$ is the dual variable and $\mathrm{I}_{B_\infty}(Y)$ is the indicator function of the $\ell_\infty$ unit norm ball.

$$\mathrm{I}_{B_\infty}(Y) = \begin{cases} 0 & \|Y\|_\infty \leq 1 \\ +\infty & otherwise \end{cases} \tag{10}$$

The min-max problem (10) can be optimal by a splitting scheme [7] as

$$X^{n+1} = arg \min_X \frac{1}{2}\|X - X^n\|_F^2 + \frac{t_1}{2}\|RFX - B\|_F^2 + t_1\lambda_1\langle \nabla X, Y^n\rangle$$

$$+ t_1\lambda_2\|X\|_* + \tau\|\varepsilon - X' - X\|_F^2 \tag{11}$$

$$X' = arg \min_{X'} t_1\lambda_1\langle \nabla X', Y^n\rangle + t_1\lambda_2\|X'\|_* + \tau\|\varepsilon - X' - X^{n+1}\|_F^2 \tag{12}$$

$$\varepsilon = arg \min_\varepsilon \|\varepsilon\|_* + \tau\|\varepsilon - X' - X^{n+1}\|_F^2 \tag{13}$$

$$Y^{n+1} = arg \max_Y \frac{1}{2}\|Y - Y^n\|_F^2 + \mathrm{I}_{B_\infty}(Y) - t_2\lambda_1\langle \nabla(2X^{n+1} + X' - X^n), Y\rangle, \tag{14}$$

Where the parameters $X^n$ and $Y^n$ are the primal and dual variables in the $n$-th iteration, respectively. $t_1$ and $t_2$ are the corresponding iteration step sizes.

(1) Firstly, we can obtain variable $X^{n+1}$ by solving subproblem (11), this challenge is reduced to a de-noising problem by simplify Eq. (7) and Ref. [4] as

$$X^{n+1} = arg \min_X \frac{1}{2}\|X - \bar{X}^n\|_F^2 + \lambda\|X\|_* \tag{15}$$

According to optimal Eq. (15), $\bar{X}^n$ can be denoted as follows

$$\bar{X}^n = X^n - \frac{t_1}{1+t_1 L}\mathcal{A}^T(\mathcal{A}X^n - B) - \frac{t_1\lambda_1}{1+t_1 L}\nabla^T Y^n + \tau(X' + \varepsilon), \; \lambda = \frac{t_1\lambda_2}{1+t_1 L} \tag{16}$$

Here the parameters $\mathcal{A} = RF$, and $L = \lambda_{max}(\mathcal{A}^T\mathcal{A})$. $\nabla^T$ is the adjoint operator of $\nabla$. Solving subproblem (15) could adapt the singular value decomposition (SVD) of variable $\bar{X}^n = Udiag(\sigma(\bar{X}^n))V^H$. Then, according to the matrix shrinkage operator [25], the optimal solution of Eq. (15) is

$$X^{n+1} = S_\lambda(\bar{X}^n) = Udiag(\bar{\sigma}_\lambda(\bar{X}^n))V^H, \bar{\sigma}_\lambda(\bar{X}^n) = \max(\sigma(\bar{X}^n) - \lambda, 0) \tag{17}$$

(2) Then, the variable $X'$ can be obtained by solving sub-problem (12), its simplest form of optimization model (9) is shown below

$$X' = \arg\min_{X'} \lambda \|X'\|_* + \frac{1}{2}\|X' - \bar{X}'\|_F^2 \qquad (18)$$

Here, the defined of parameter $\bar{X}'$ can be listed as follows:

$$\bar{X}' = X^{n+1} - \frac{t_1 \lambda_1}{1 + t_1 L} \nabla^T Y^n + \tau \varepsilon \qquad (19)$$

Therefore, the solution of Eq. (18) can be obtained by the matrix shrinkage operator as

$$X' = S_\lambda(\bar{X}') = U diag(\bar{\sigma}_\lambda(\bar{X}'))V^H, \bar{\sigma}_\lambda(\bar{X}') = \max(\sigma(\bar{X}') - \lambda, 0) \qquad (20)$$

(3) Solving sub-problem (13) also requires the use of the matrix shrinkage operator, and the optimal result can be solved as follows

$$\varepsilon = S_\lambda(\varepsilon') = U diag(\bar{\sigma}_\lambda(\varepsilon'))V^H \qquad (21)$$

$$\varepsilon' = X' - X^{n+1} \qquad (22)$$

(4) Then we consider the $Y$ subproblem in Eq. (14)

$$Y^{n+1} = \arg\max_Y \frac{1}{2}\|Y - \bar{Y}^n\|_F^2 + I_{B_\infty}(Y) \qquad (23)$$

**Lemma 1.** According to Ref. [23], the forward difference operator of $X$ denotes as $\nabla X$, and $\nabla X = (P, Q)$, where $P \in \mathbb{C}^{(m-1) \times n}$ and $Q \in \mathbb{C}^{n \times (m-1)}$. $P_{i,j} = x_{i,j} - x_{i+1,j}$, $Q_{i,j} = x_{i,j} - x_{i,j+1}$. We can get

$$\langle \nabla X, Y \rangle = \langle X, \nabla^H Y \rangle \qquad (24)$$

$$(\nabla^H Y)_{i,j} = (\nabla^H (P, Q))_{i,j} = P_{i,j} + Q_{i,j} - P_{i-1,j} - Q_{i,j-1} \qquad (25)$$

Where the dual variable $Y$ is also constructed by the matrix pair $(P, Q)$, and the adjoint operator of $\nabla$ denoted by $\nabla^H$. Therefore, $\bar{Y}^n$ can defined as

$$\bar{Y}^n = Y^n + t_2 \lambda_1 \nabla(2X^{n+1} + X' - X^n) \qquad (26)$$

The solution of objection function (23) is non-convex, it cannot be optimized by direct derivation. Like sub-problem (23) can be obtained by the Euclidean projection of $\bar{Y}^n$ onto $\ell_\infty$ unit norm ball, so it can be evaluated by

$$Y^{n+1} = sgn(\bar{Y}^n) \cdot min\ (|\bar{Y}^n|, 1) \tag{27}$$

Where all the operations in Eq. (27) are the element-wise, and $sgn(x)$ is the sign function,

$$sgn(x) = \begin{cases} 1 & x > 0 \\ 0 & x < 0 \end{cases} \tag{28}$$

Through solving and analyzing the above model, we clearly obtained the optimization process of variables $X, X', \varepsilon$ and $Y$, and summarized the solving steps of variables in **Algorithm 1**

---
**Algorithm 1** RDLEDM.

---
**Input:** $\mathcal{A} = RF, B, \lambda_1, \lambda_2$

**initialization:** $X_0, Y_0, X', \varepsilon, \lambda = t_1\lambda_2/1 + t_1 L$

**while** not converged **do**

  1) Compute: $\bar{X}^n = X^n - \frac{t_1}{1+t_1 L}\mathcal{A}^T(\mathcal{A}X^n - B) - \frac{t_1\lambda_1}{1+t_1 L}\nabla^T Y^n + \tau(X' + \varepsilon)$ in Eq (11)

  2) Acquire the solution to variable $X^{n+1}$ by evaluating Matrix Shrinkage Operator: $X^{n+1} = S_\lambda(\bar{X}^n)$ in Eq. (13)

  3) Compute: $\bar{X}' = X^{n+1} - \frac{t_1\lambda_1}{1+t_1 L}\nabla^T Y^n + \tau\varepsilon$ in Eq. (19)

  4) Acquire the solution to variable $X'$ by SVD $X' = S_\lambda(\bar{X}')$ in Eq. (20)

  5) Acquire the solution to variable $\varepsilon$ by SVD: $\varepsilon = S_\lambda(X' - X^{n+1})$ in Eqs. (21) and (22)

  6) Compute: $\bar{Y}^n = Y^n + t_2\lambda_1\nabla(2X^{n+1} + X' - X^n)$ in Eq. (26)

  7) Acquire the solution to variable $Y^{n+1}$ by projecting $\bar{Y}^n$ onto $\ell_\infty$ unit ball: $Y^{n+1} = sgn(\bar{Y}^n) \cdot min\ (|\bar{Y}^n|, 1)$

**end while**

**output:** $X$

---

# 5 Experiments

In this section, we first test the validity of the proposed method on three noisy dynamic MR sequences. To simulate the noisy dynamic MR images in k-space, we used the in-vivo breath-hold cardiac perfusion dynamic MRI data (192×192×40) [23], cine data (128×128×20) **[**26**]** and cerebral perfusion data (128×128×60) of reference [27]. Then, gaussian white noise of standard derivation σ = 0.05 is added to k-space data, as shown in Eq. (2). We apply the most practical 2D Random mask, Pseudo Radial mask, and Cartesian mask with different sampling ratio as the under-sampled mask in our experiments [28]. As shown in Fig. 2 below, 2D Random sampling, Pseudo Radial sampling, and Cartesian sampling at 25% sampling rate are listed respectively.

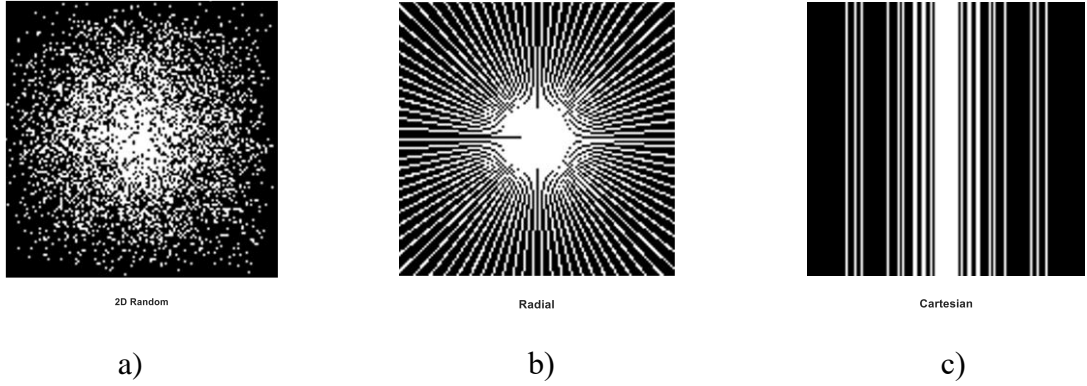

| 2D Random | Radial | Cartesian |
| a) | b) | c) |

Figure 2: (a)-(c) are the 2D Random mask, the Pseudo Radial mask, and the Cartesian mask at 25% sampling rate, respectively.

To explain the performance of our proposed method, five authoritative comparison methods were referred to in the experimental stage, including BCS [29], KTSLR [30], SDR [31], FTVNNR [23], and SRTPCA [13]. The source code for these methods was downloaded from each author's web site, and we used their default Settings or tuned their results to the best of our experiments with five state-of-the-art methods. For quantitative evaluation analysis, PSNR and RMSE are adopted as the metric.

$$PSNR(x,\hat{x}) = 10 * log10 \frac{255^2}{\|x-\hat{x}\|_2^2/N} \tag{29}$$

$$RMSE(x,\hat{x}) = \sqrt{\frac{1}{mn}\sum_{i=1}^{m}\sum_{j=1}^{n}(x_{ij}-\hat{x}_{ij})^2} \tag{30}$$

Where $x$ and $\hat{x}$ are fully-sampled and reconstructed images, respectively; the higher PSNR and lower RMSE indicate the reconstructed $\hat{x}$ is more closer to original $x$.

All experiments are conducted using MATLAB 2018b on a desktop with AMD Ryzen 9 5900HX 3.3GHz CPU and 32.0 GB RAM.

**5.1 Comparison on Visual Quality**

To prove the superiority of the proposed method, the reconstruction quality is evaluated by visual inspection. The first experiment, we adopt the 25% Cartesian sampling pattern to generate the k-space data of the in-vivo breath-hold cardiac perfusion dynamic MRI sequences. Fig. 3 shows the reconstruction results of cardiac perfusion dynamic 40th MRI frame, which are produced by different methods. From Fig. 3, it can be clearly found that the reconstruction results of comparison methods BCS and KTSLR are relatively not perfect, and the processing performance of these two methods for noise of this class is not so perfect. At the same time, the SDR produces obvious noise-like artifacts in the background and smooth areas, and the restored edges and textures are severely degraded. The second-best reconstruction method FTVNNR is shown in Fig. 3(g), which reconstruction effect is very outstanding for other comparison methods, both in terms of boundary and texture information, which highlights the effect of TV and NN regularizations on the superior performance of this method to suppress such noise. Compared with our proposed method, FTVNNR is smoother in surface texture reconstruction, resulting in a slight weakness in detail compared with our method. However, the proposed method greatly suppresses artifacts in background and produces finer details with a high resolution, demonstrated in Fig. 3(h).

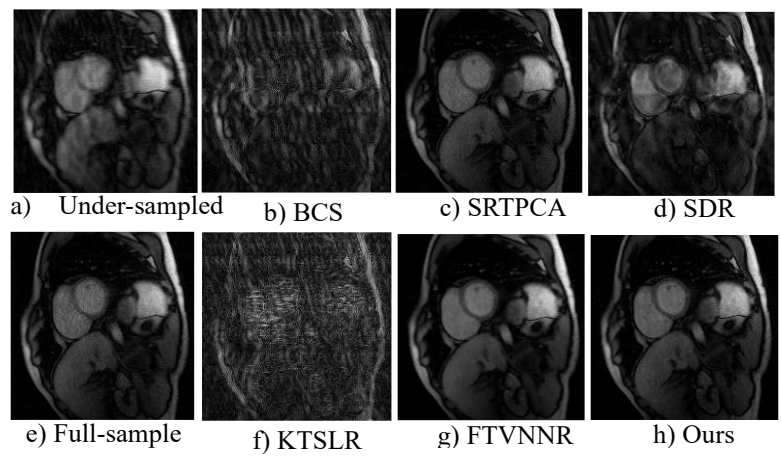

Figure 3: Reconstruction of the in-vivo breath-hold cardiac perfusion dynamic 40th MRI frame from 25% sample rate cartesian mask k-space. (a) The under-sampled image by 25% sample rate cartesian k-space; (b)–(d) and (f)-(h) reconstructed images using BCS, SRTPCA, SDR, KTSLR, and FTVNNR and our proposed method, respectively; (e) The full-sampled image.

Fig. 4 presents the reconstruction results of the cerebral perfusion MRI sequences (60th frame) that produced by different methods and the 25% Cartesian sampling pattern. The cerebral Perfusion MRI sequences are data that can better verify the effectiveness of the algorithm in texture and boundary detail reconstruction performance. The method BCS framework used to recover dynamic MRI sequences from under-sampled measurements showed the worst visual effect, and neither details nor boundary areas could be reconstructed well to meet the needs of reality. This is because when low-rank sparse decomposition is carried out, only the smoothness of coefficient factors is considered, and the effect of texture and boundary factors of details is not considered. The reconstruction effect of SDR is second only to that of BCS, The SDR model further uniformly describes the sparse plus low-rank characteristics, again without specific consideration of details. Although The SRTPCA and KTSLR use TV regularization and NN regularization respectively to consider part of texture and boundary information, they fail to fully reconstruct more details. The second-best visual reconstruction method FTVNNR reduces artifacts in different degrees and recovers more details than other comparative methods. However, the proposed algorithm greatly suppresses artifacts in background and produces finer details with a high resolution.

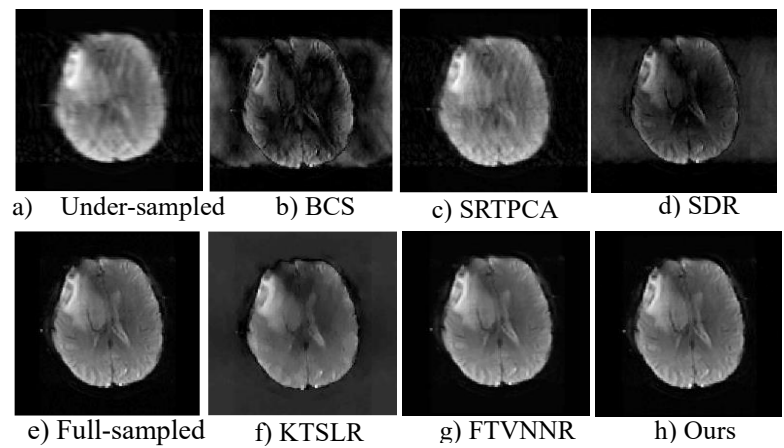

a) Under-sampled   b) BCS   c) SRTPCA   d) SDR

e) Full-sampled   f) KTSLR   g) FTVNNR   h) Ours

Figure 4: Reconstruction of the cerebral perfusion data dynamic 60th MRI frame from 25% sample rate

cartesian mask k-space. (a) The under-sampled image by 25% sample rate cartesian k-space; (b)–(d) and (f)-(h) reconstructed images using BCS, SRTPCA, SDR, KTSLR, and FTVNNR and our proposed method, respectively; (e) The full-sampled image.

Similarly, Fig. 5 shows the visualized reconstruction effect of the Cine MRI sequences (20th), which is also processed by cartesian mask under-sampled with 25% sampling rate. Its actual under-sampled visualization is shown in Fig. 5 (a). Some boundaries and textures on the image surface of each frame become very blurred. Among them, the visual reconstruction effect of the comparison methods SDR and BCS is not so perfect, and the large number of boundaries and textures are not well recovered from the under-sampled noise data. On the other hand, the visualized reconstruction effect of SRTPCA and KTSLR is obviously better than that of SDR and BCS mentioned above, and the MRI sequences reconstructed by SRTPCA and KTSLR have advantages over SDR and BCS in boundary. However, there are still shortcomings in detail texture. It can be clearly seen from Fig. 5 (c) and (f) that the surface texture of MRI sequence reconstructed by these two methods is a little fuzzy, which seriously affects the visualization effect. As the second FTVNNR with good visual effect, although it looks not much different from the full-sampled data, we can still find that the texture FTVNNR in some areas has deviations from the method we proposed. Compared with FTVNNR, the visual reconstruction effect of the proposed method is more distinct in the boundary part, and the texture details of the surface of some areas are more localized, which is not as smooth as the average smoothness of FTVNNR's large area.

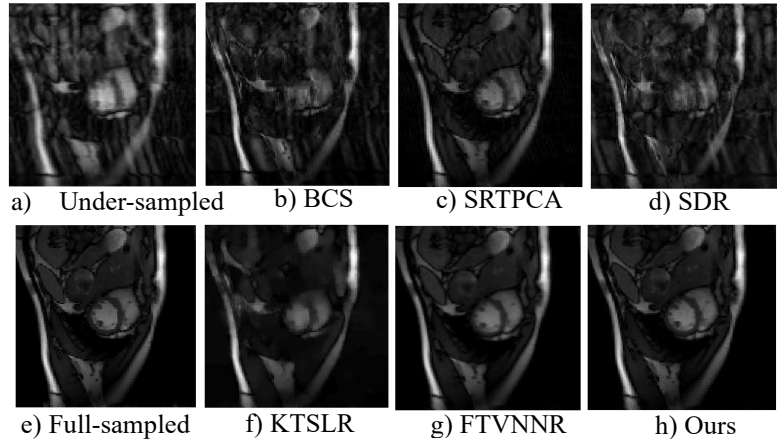

a) Under-sampled    b) BCS    c) SRTPCA    d) SDR

e) Full-sampled    f) KTSLR    g) FTVNNR    h) Ours

Figure 5: Reconstruction of the cine data dynamic 20th MRI frame from 25% sample rate cartesian mask k-space. (a) The under-sampled image by 25% sample rate cartesian k-space; (b)–(d) and (f)-(h) reconstructed images using BCS, SRTPCA, SDR, KTSLR, and FTVNNR and our proposed method, respectively; (e) The full-sampled image.

### 5.2 The Quantitative Evaluation Analysis

We further verify the effectiveness of our proposed method in dynamic MRI sequence reconstruction. The experimental analysis in this subsection mainly focuses on two quantitative indicators PSNR and RMSE, compared with several other state-of-the art dynamic MRI sequence reconstruction methods. As shown in Tables 1 and 2 below, the cardiac Perfusion MRI data, Cine data and cerebral Perfusion data are respectively presented to evaluate the reconstruction quality of Cartesian mask at 25% sampling rate.

Table 1 shows the comparative data of all methods of quantitative evaluation index PSNR in dynamic MRI sequence reconstruction. From the table, it is not difficult to find that compared with other methods, the proposed method has significantly improved in the cardiac Perfusion MRI data, Cine data and cerebral Perfusion data. Among them, Cine data in dynamic MRI sequence increased relatively significantly, about 2.5dB. It's seen that our proposed method has advantages in the reconstruction of some boundaries. The reconstructed quality of the data set, Cerebral Perfusion, was also improved compared to the second-best comparison method, FTVNNR, which was about 0.45dB. This dataset is a data set focusing on surface detail texture reconstruction, which indicates that the method proposed by us does not lag-behind in surface texture reconstruction on details. Similarly, the cardiac Perfusion dataset also achieved good reconstruction results, with an increase of about 1.02 dB. This data set has a larger dimension than the other two dynamic MRI sequences.

Table 2 shows the RMSE quality assessment results of the three dynamic MRI sequences in all methods. It can be seen from the table that the RMSE values of BCS are all large in these three dynamic MRI sequences, which also indicates that the quality of reconstruction is very inferior. Compared with the method proposed by us and BCS, the double TV and double NN regression proposed by us have great advantages, which are -0.1395, -0.2375 and -0.5298 respectively. Compared with FTVNNR, which is the second-best comparison method, our algorithm is also higher, which are -0.0036, -0.0007 and -0.0003 respectively. The analysis of these quantitative indexes shows the superiority of our method.

Table 1: The PSNR (dB) results under 25% sample rate cartesian k-space with different methods (the best results are in bold).

| Dataset\Methods | BCS | SRTPCA | SDR | KTSLR | FTVNNR | Ours |
|---|---|---|---|---|---|---|
| Perfusion (192×192×40) | 18.91 | 31.21 | 20.56 | 16.85 | 33.53 | **34.55** |
| Cine (128×128×20) | 19.97 | 25.81 | 20.43 | 24.63 | 33.41 | **36.18** |
| Cerebral(128×128×60) | 16.43 | 30.45 | 18.19 | 18.98 | 41.22 | **41.67** |

Table 2: The RMSE results under 25% sample rate cartesian k-space with different methods (the best results are in bold).

| Dataset\Methods | BCS | SRTPCA | SDR | KTSLR | FTVNNR | Ours |
|---|---|---|---|---|---|---|
| Perfusion (192×192×40) | 0.1461 | 0.0119 | 0.1929 | 0.4029 | 0.0102 | **0.0066** |
| Cine (128×128×20) | 0.2460 | 0.0710 | 0.1718 | 0.0664 | 0.0092 | **0.0085** |
| Cerebral(128×128×60) | 0.5319 | 0.0268 | 0.3899 | 0.2949 | 0.0025 | **0.0021** |

## 5.3 Convergence of the Proposed Algorithm

In Fig. 6, we use three dynamic MRI data sets sampled at 25% cartesian mask to demonstrate the experimental convergence performance of the proposed method. This is the actual convergence graph obtained when the setting parameters of each data set are optimal. The ordinate of this graph is the MRI sequence correlation error value reconstructed in each iteration, and the other correlation error values are defined as $RE = \|x^{t+1} - x^t\|_F^2 / \|x^t\|_F^2$, which are used to quantitatively measure the accuracy performance in each iteration. It can be seen from Fig. 6 that RE decreases sharply from the beginning of iteration. When the number of iterations is greater than 50, RE of both data sets becomes smaller and tends to be more stable later. This shows that the performance of our proposed method has good universality and stability.

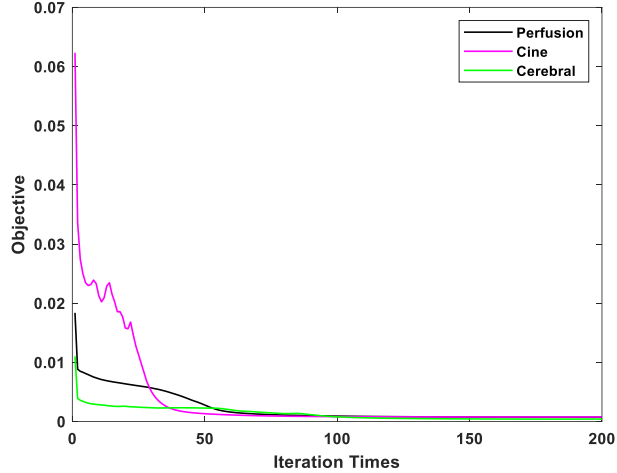

Figure 6: The convergence results of different dynamics MRI sequences Perfusion, Cine and Cerebral under 25% sample rate at cartesian k-space.

## 5.4 Reconstruction with Different Sampling Patterns

For practical consideration, we conduct experiments on three dynamic MRI sequences sampled data polluted by different under-sampled levels to evaluate the performance of the proposed method. In the experiments, we recorded PSNR and RMSE results of k-space data reconstruction under

Cartesion mask, radial Mask and 2D random mask with 20%-65% sampling rate. Fig. 7-9 clearly shows that the quantitative index PSNR is directly proportional to the sampling rate of the sample, and PSNR increases with the increase of the sampling rate. Both Cartesion and Radial and 2D Random samples follow this proportional relationship. On the contrary, the quantitative index RMSE is inversely proportional to the sampling rate of the sample, and RMSE decreases with the increase of the sampling rate. At the same time, this inverse relationship is also established for the three under-sampled pattern. These two proportional relations should be exactly the relationship between Eq. (21) and Eq. (22). It completely verifies the rationality of our proposed algorithm theory. It is worth noting that the size relationship between PSNR and RMSE among the three sampling modes can be seen from Figure 7-9. The reconstruction effect of the dynamic MRI sequence of the sampling mode is better than that of the Radial mode and the Cartesion mode under 2D Random pattern. At the same time, the maximum PSNR and minimum RMSE values were obtained in the three samples by using dynamic MRI sequence cerebral Perfusion data. Particularly, Fig. 8, the Radial mode, the hierarchical relationship of the three dynamic MRI sequences is very clear, and the quantified and differentiated effects of the three reconstruction effects are obvious.

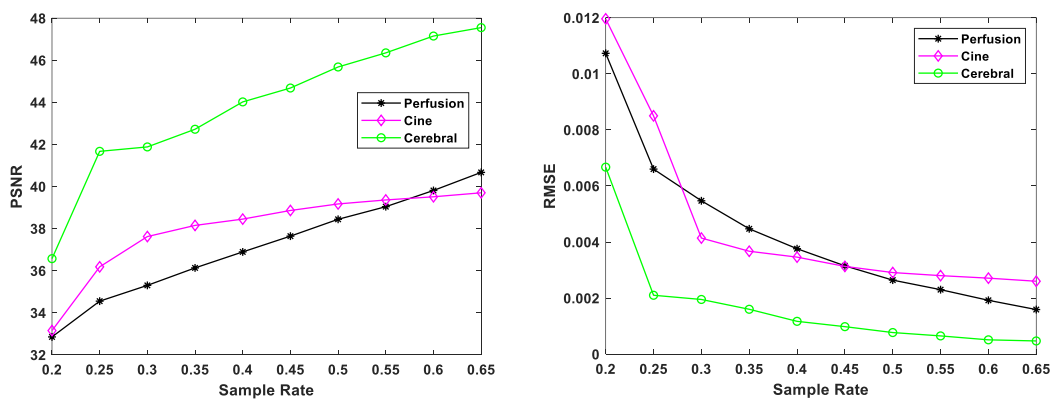

Figure 7: The quantitative evaluation metric PSNR and RMSE results of dynamic MRI sequences the cardiac Perfusion, Cine and cerebral Perfusion under the different sample rate Cartesian mask, and left

is quantitative evaluation metric PSNR, right is quantitative evaluation metric RMSE.

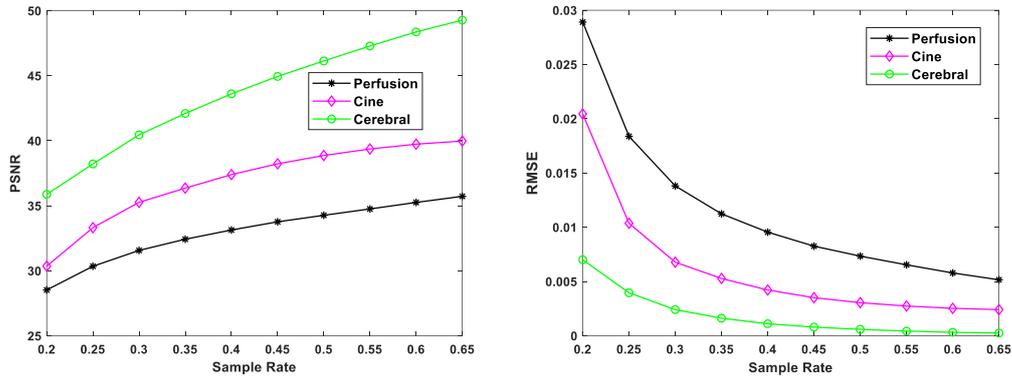

Figure 8: The quantitative evaluation metric PSNR and RMSE results of dynamic MRI sequences the cardiac Perfusion, Cine and cerebral Perfusion under the different sample rate Radial mask, and left is quantitative evaluation metric PSNR, right is quantitative evaluation metric RMSE.

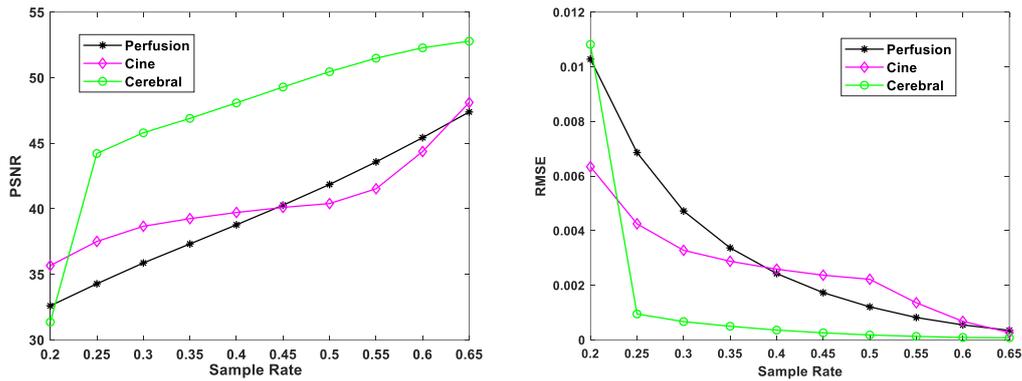

Figure 8: The quantitative evaluation metric PSNR and RMSE results of dynamic MRI sequences the cardiac Perfusion, Cine and cerebral Perfusion under the different sample rate 2D Random mask, and left is quantitative evaluation metric PSNR, right is quantitative evaluation metric RMSE.

# 6 Conclusion

This paper proposed a novel robust sparse decomposition k-space reconstruction framework for the application of dynamic MRI sequences. The proposed method employed two kinds of transform to establish a mixed noise regularization model and improve k-space MRI sequences reconstruction quality, which can exploit the advantage of the DFT transform and linear transform domain, simultaneously. By introducing double TV regularization and double NN regularization, which preserve the edge structural details in non-smooth regions and piecewise-smooth information of image in smooth regions, separately. The solution of the proposed method was given by the fast iterative algorithm. Various experimental results demonstrate that the proposed RDLEDM can achieve the superior performance in detail clarity and noise suppression from the objective and subjective visual evaluation.

## References

[1] Haldar J. P., Hernando D., Liang Z. P. Compressed-sensing MRI with random encoding[J]. IEEE transactions on Medical Imaging, 2010, 30(4): 893-903.

[2] Li S., Da Xu. L., Wang X. Compressed sensing signal and data acquisition in wireless sensor networks and internet of things[J]. IEEE Transactions on Industrial Informatics, 2012, 9(4): 2177-2186.

[3] Bernstein B., Liu S., Papadaniil C., et al. Sparse recovery beyond compressed sensing: Separable nonlinear inverse problems[J]. IEEE transactions on information theory, 2020, 66(9): 5904-5926.

[4] J. Tan, X. Zhang, C. Qing and X. Xu, "Fourier Domain Robust Denoising Decomposition and Adaptive Patch MRI Reconstruction," in IEEE Transactions on Neural Networks and Learning Systems, doi: 10.1109/TNNLS.2022.3222394.

[5] Christodoulou A. G., Lingala S. G. Accelerated dynamic magnetic resonance imaging using learned representations: a new frontier in biomedical imaging[J]. IEEE Signal Processing Magazine, 2020, 37(1): 83-93.

[6] Ma S., Yin W., Zhang Y., et al. An efficient algorithm for compressed MR imaging using total

variation and wavelets[C]//2008 IEEE Conference on Computer Vision and Pattern Recognition. IEEE, 2008: 1-8.

[7] Blunck Y., Kolbe S. C., Moffat B. A., et al. Compressed sensing effects on quantitative analysis of undersampled human brain sodium MRI[J]. Magnetic Resonance in Medicine, 2020, 83(3): 1025-1033.

[8] Guruprasad S., Bharathi S. H., Delvi D. A. R. Effective compressed sensing MRI reconstruction via hybrid GSGWO algorithm[J]. Journal of Visual Communication and Image Representation, 2021, 80: 103274.

[9] Ulas C., Gómez P. A., Sperl J I, et al. Spatio-temporal MRI reconstruction by enforcing local and global regularity via dynamic total variation and nuclear norm minimization[C]//2016 IEEE 13th International Symposium on Biomedical Imaging (ISBI). IEEE, 2016: 306-309.

[10] Yang X., Mei Y., Hu X., et al. Compressed sensing MRI by integrating deep denoiser and weighted Schatten p-norm minimization[J]. IEEE Signal Processing Letters, 2021.

[11] Chen Z., Huang C., Lin S., A new sparse representation framework for compressed sensing MRI[J]. Knowledge-Based Systems, 2020, 188: 104969.

[12] Mubarak M., Thomas T. J., Rani J. S., et al. Higher order Dictionary Learning for Compressed Sensing based Dynamic MRI reconstruction[C]//BMVC. 2019: 190.

[13] Liu Y., Liu T., Liu J., et al. Smooth robust tensor principal component analysis for compressed sensing of dynamic MRI[J]. Pattern Recognition, 2020, 102: 107252.

[14] Chang M., Li Q., Feng H., et al. Spatial-adaptive network for single image denoising[C]//European Conference on Computer Vision. Springer, Cham, 2020: 171-187.

[15] Li A., Zhao S., Ma X., et al. Short-term and long-term context aggregation network for video

inpainting[C]//European Conference on Computer Vision. Springer, Cham, 2020: 728-743.

[16] Liu S. L., Guo H. X., Pan H., et al. Deep implicit moving least-squares functions for 3d reconstruction[C]//Proceedings of the IEEE/CVF Conference on Computer Vision and Pattern Recognition. 2021: 1788-1797.

[17] Cole E. K., Ong F., Vasanawala S. S., et al. Fast unsupervised mri reconstruction without fully-sampled ground truth data using generative adversarial networks[C]//Proceedings of the IEEE/CVF International Conference on Computer Vision. 2021: 3988-3997.

[18] Fan Z., Sun L., Ding X., et al. A segmentation-aware deep fusion network for compressed sensing mri[C]//Proceedings of the European Conference on Computer Vision (ECCV). 2018: 55-70.

[19] Yoo J., Jin K. H., Gupta H., et al. Time-dependent deep image prior for dynamic MRI[J]. IEEE Transactions on Medical Imaging, 2021, 40(12): 3337-3348.

[20] Fabian Z, Heckel R, Soltanolkotabi M. Data augmentation for deep learning based accelerated MRI reconstruction with limited data[C]//International Conference on Machine Learning. PMLR, 2021: 3057-3067.

[21] Zou Q., Ahmed A. H., Nagpal P., et al. Dynamic imaging using a deep generative SToRM (Gen-SToRM) model[J]. IEEE transactions on medical imaging, 2021, 40(11): 3102-3112.

[22] Candès E. J., Wakin M. B., An introduction to compressive sampling[J]. IEEE signal processing magazine, 2008, 25(2): 21-30.

[23] Yao J., Xu Z., Huang X., et al. An efficient algorithm for dynamic MRI using low-rank and total variation regularizations[J]. Medical image analysis, 2018, 44: 14-27.

[24] Boyd S., Boyd S. P., Vandenberghe L., Convex optimization[M]. Cambridge university press,

2004.

[25] Ma S., Goldfarb D., Chen L., Fixed point and Bregman iterative methods for matrix rank minimization[J]. Mathematical Programming, 2011, 128(1): 321-353.

[26] Liu Y., Yi Z., Zhao Y., et al. Calibrationless parallel imaging reconstruction for multislice MR data using low-rank tensor completion[J]. Magnetic Resonance in Medicine, 2021, 85(2): 897-911.

[27] Lin C. Y., Fessler J. A. Efficient dynamic parallel MRI reconstruction for the low-rank plus sparse model[J]. IEEE transactions on computational imaging, 2018, 5(1): 17-26.

[28] Cao J., Liu S., Liu H., et al. CS-MRI reconstruction based on analysis dictionary learning and manifold structure regularization[J]. Neural Networks, 2020, 123: 217-233.

[29] Lingala S. G., Jacob M., Blind compressive sensing dynamic MRI[J]. IEEE transactions on medical imaging, 2013, 32(6): 1132-1145.

[30] Lingala S. G., Hu Y., DiBella E., et al. Accelerated dynamic MRI exploiting sparsity and low-rank structure: kt SLR[J]. IEEE transactions on medical imaging, 2011, 30(5): 1042-1054.

[31] Liu Q., Wang S., Liang D., Sparse and dense hybrid representation via subspace modeling for dynamic MRI[J]. Computerized Medical Imaging and Graphics, 2017, 56: 24-37.